# Is dark matter in spiral galaxies cold gas?
## I. Observational constraints and dynamical clues about galaxy evolution

**Daniel Pfenniger**[1], **Françoise Combes**[2], **Louis Martinet**[1]

[1] Observatoire de Genève, CH-1290 Sauverny, Switzerland
[2] DEMIRM, Observatoire de Meudon, 92 190 Meudon, France

**Abstract.** Based on dynamical constraints about the Hubble sequence evolution, observational data and a number of "conspiracies", we propose that the dark matter around spiral galaxies is in the form of cold gas, essentially in molecular form and rotationally supported.

The existence of a much larger amount of cold gas in the outer discs of spirals is in accordance with dynamical studies taking into account the bar phenomenon and the secular growth of bulges, leading to the general conclusion that spiral galaxies have to evolve along the Hubble sequence from Sd to Sa over a couple of Gyr's. If so, the varying $M/L$ ratio along the Hubble sequence suggests that dark matter is transformed into stars, i.e. dark matter should be in a sufficiently diluted form of gas.

This hypothesis sheds light on several puzzles: 1) the conspiracy of the flat rotation curves, 2) the estimated short gas consumption time of spiral galaxies, leading to the "gas consumption problem", 3) the constant ratio of dark matter to HI mass in the outer spiral discs, 4) the larger amount of visible gas in interacting galaxies with respect to isolated ones, giving rise to gigantic starbursts, 5) the high frequency of absorption lines in front of quasars, together with the large sizes of the absorbing medium around galaxies.

Several problems raised by a completely gaseous dark matter in the outer spiral discs are discussed. In particular the stability and self-consistency of cold gaseous discs is shown to be a less severe problem than commonly believed.

## 1. Introduction

Numerous assumptions have been made on the nature of dark matter, from modest extrapolation of well known objects, such as brown dwarfs, to a prolific variety of much more speculative hypotheses, such as WIMPS or modified gravity. Since dark matter seems to be required in different amounts over very different scales, it is physically reasonable to expect that different kinds of explanations might be necessary along the cosmic hierarchy (Rees 1987). In some way the amount of dark matter is a measure of our ignorance, or of our lack of confidence in existing theories[1], about the phenomena acting at each scale. A priori we have no reason to expect that our ignorance should be restricted to a single type of matter or physical effect.

The present discussion is directed principally towards the problem of dark matter at the level of individual disc galaxies, for which it is the clearest. Therefore, a review about the main steps of the evolution of the ideas about dark matter inside these objects is useful.

*1.1. Dynamics in a historical perspective*

As well known, since Zwicky's early work (1933) dynamical evidence existed for large amounts of undetected matter in clusters of galaxies. Only forty years later, studies of field spiral kinematics by HI 21 cm observations (Rogstad & Shostak 1972; Roberts & Rots 1973) gave a new impetus to this problem at the scale of galaxies: flat rotation curves suggested that the presence of hidden forms of matter is a property of isolated spirals as well.

In the early seventies two kinds of results appeared in conjunction: whereas the first flat rotation curves were published, seminal papers by Miller & Prendergast (1968) and Hohl & Hockney (1969) presented $N$-body simulations of disc galaxies in view of explaining the formation and maintenance of spiral structures in discs. At that time, the galactic astronomers were focused on the linear density wave theory (Lin & Shu 1964; Goldreich & Lynden-Bell 1965) concerning tightly wound spirals. The central problem for people working on $N$-body numerical experiments was to suppress these strong bars which form

*Send offprint requests to*: D. Pfenniger

---

[1] For example most of the mass in the solar system, although hardly directly observable, is not considered as "dark" because the gravity of its bodies is consistent with the theories of their mass distribution.

spontaneously in the simulations, in order to obtain the well defined grand design spiral structure, "so often observed" and "well explained by the density wave theory". In fact, in the seventies, bars were not much studied and practically not observed, with the noticeable exception of de Vaucouleurs who insisted many times on their importance in the spiral classification. A further motivation to ignore them was that bars could hardly be included in the density wave theory. Essentially, two solutions were often considered to prevent the formation of bars: hot discs or massive halos (Ostriker & Peebles 1973; Einasto et al. 1974). But in the seventies no data could indicate that the inner parts of the galactic discs were hot enough. Discs were considered as consisting of cold populations by extrapolation of the solar neighbourhood data. Now, more recent observations indicate that the velocity dispersions in inner parts of typical Sbc galaxies can be of the order of $100 \, {\rm km \, s^{-1}}$ (e.g. Lewis & Freeman 1989; Bottema 1993), showing that discs can be hot there.

In the last decade, theoretical arguments, as well as new numerical experiments and deductions from the observations of the luminosity profile of discs, have contributed to modify the point of view about the rôle played by halos on global stability. The stabilisation of discs against non-axisymmetric perturbations, if ensured by halos, should only depend on the halo mass located in the inner region of the disc, where a bar usually develops; it should not depend on the halo mass in the outer regions, where the rotation curve remains flat. Kalnajs (1987) disturbed minds by claiming 1) that the "luminous" rotation curves, calculated from the observed exponential stellar disc luminosity profiles, would fit very well the observed rotation curves in the optical region of galaxies, and, 2) that, as a consequence, dark halos would be useless to stabilise against a bar.

Today, the global stability in the discs against bar-like perturbations is no longer the acute problem that it used to be in the seventies: at least two thirds of observed spirals are recognised to be barred or to present an oval structure in the inner regions (de Vaucouleurs 1963). The remaining third include edge-on and dusty galaxies in which a bar can hardly be detected. In many cases, IR photometry, less affected by dust, reveals barred structures invisible in the B band; the most famous case being the Milky Way bar (Blitz et al. 1993). Further on, extensive numerical simulations by Athanassoula & Sellwood (1986) have shown that the growth of a bar may be prevented if the central part of the disc is hot enough. Finally, recent numerical experiments indicate how bars can be destroyed (Pfenniger & Norman 1990; Pfenniger 1991a; Friedli & Benz 1993). Such dynamical constraints have implications on dark matter as explained in more detail in Section 2.

### 1.2. HI rotation curves, maximum disc

HI 21 cm observations of rotation curves have to be joined to the previous considerations. Spiral HI rotation curves, which extend typically up to $2-3\,R_{25}$, are well known to stay in general sufficiently flat over this range for requiring an important presence of mass in the outer parts of spirals. Some authors tried to solve the question of the contribution of various components (disc, bulge, halo) to the observed rotation curves produced by HI data (see e.g. Sancisi & van Albada 1987). Among the proposed solutions, the "maximum disc" solution consists in maximising the contribution of the luminous matter to the observed rotation curve. This yields to a conservative lower limit for the amount of dark matter and an estimate of the $M/L$ ratio for the disc. The maximum disc solution received recently a strong support from an extensive work on kinematical data of $\sim 500$ galaxies; it reproduces well the fine wiggles of the light, confirming that the fraction of dark halo within the stellar optical parts of the galaxies must be negligible (Freeman 1992). Within the error margins, a small amount of dark matter in the optical disc is consistent with the estimates of the solar neighbourhood dark matter (Oort 1932; Bienaymé et al. 1987; Kuijken & Gilmore 1989; Bahcall et al. 1992) which represents between 0 and 50% of the local mass, though both extreme values are controversial.

### 1.3. Conspiracy

*The only really obvious dark matter problem subsisting in spirals concerns the outer regions beyond the optical disc.* The fact that the disc plus the bulge on one hand, and the dark halo on the other hand, bring an essential and about equal contribution to the flat rotation curve in two distinct regions, namely in the inner luminous region and in the outer 21 cm emitting HI region, has suggested a physical coupling between these components, also called disc-halo conspiracy (Bahcall & Casertano 1985; van Albada & Sancisi 1986). No convincing explanation of it has yet been given.

Recently new HI observations have restricted the range of the conspiracy: Casertano & van Gorkom (1991) published HI rotation curves characterised by a large decrease between 1 and $3\,R_{25}$ and have found a clear correlation between the peak circular velocity, its central brightness and the slope of the rotation curve in the outer parts of the discs. This result leads these authors to suggest that the ratio of the dark to luminous matter might be the critical parameter controlling the Hubble sequence. It is crucial to determine if really late-type galaxies contain more dark matter than early-type ones, or if the importance of dark matter decreases with luminosity. These questions introduced by Tinsley (1981) have been recently discussed again (see for example Salucci et al. 1991 and references therein).

### 1.4. Bosma HI/dark matter ratio

A curious coincidence (Bosma 1981), until now unexplained, has been recently recalled by Freeman (1992, and references therein) and better documented by Broeils (1992): the surface density ratio of dark matter and HI gas in a sample of galaxies remains constant outside the optical disc, around $10-30$. If dark matter tends to follow HI with a constant ratio of about 20, we have the very curious situation that dark matter is concomitant to HI radially, yet dark matter does not prevent HI to flare out the plane just beyond the optical disc. The flaring shows at least that dark matter cannot be much flatter than HI. On the other hand, attempts to constrain the flattening of dark matter by the thickness of the flaring HI layer are either inconclusive (Kundić et al. 1993), or indicate a flattened distribution, possibly as flat as HI (Rob Olling, private communication, and Sect. 4.1). Certainly the simplest assumption is that if dark matter is proportional to HI radially, then it should also be proportional vertically.

*1.5. Overview*

To solve the previously summarised problems and conspiracies, we propose in this and a companion paper (Pfenniger & Combes 1994, Paper II) the following hypothesis: dark matter in spiral galaxies is made essentially of cold molecular gas, in a mostly rotation supported disc. In the numerous conferences dedicated to the dark matter problem, extremely little critical attention has been devoted to this possibility.

In Section 2 we review several new constraints that advances in the large scale dynamics of galaxies have furnished. In particular the general conclusion that can be drawn from numerical dynamical studies is that spirals are unsteady objects and they have to evolve secularly along the Hubble sequence in the sense Sd $\rightarrow$ S0, either by smooth evolution driven by a bar, or more dramatically by galaxy-galaxy collisions. Since the ratio of dark matter to stellar mass decreases along the Hubble sequence, from Sd to Sa, we can infer that stars should arise from dark matter, i.e. dark matter should be in a sufficiently diluted form of hydrogen, excluding Jupiters, brown and white dwarfs, and black-holes.

In Section 3 we review the different observational constraints on the possible forms of hydrogen. A massive amount of hot or warm form of hydrogen can be ruled out, essentially because hot and warm gas already fills most of the interstellar volume at a much too low density to contribute to the mass in an appreciable amount. On the contrary, although cold gas fills a small part of the volume, its density is large enough to encompass most of the ISM mass. Moreover, this condensed structure explains in part the invisible character of the medium, its low cross-section for absorption studies, and its transparency to external radiation. Even if the medium is essentially molecular, the CO molecule can hardly serve as a tracer, because of the low metallicity of this quasi-primordial gas. A fraction could also be in atomic form, but HI emission cannot be detected in a medium at 3 K. Absorption studies of the outer parts of galaxies could probe this medium.

Based on the fact that cold gas is observed to be fractal over several decades of length and density, a more specific discussion of the possible state of cold gas is presented in Paper II. It turns out that both the problem of mass underestimate in HI discs and the problem of star non-formation in outer discs is closely linked to the fractal structure. The physical state of this gas must be high density and cold temperature. Since no significant heating sources in the outer discs presumably exist, we can assume that the gas is bathing in the cosmological background, and that its temperature is about 3 K. In these nearly isothermal conditions, clouds can fragment until they reach small clump units, where the cooling time becomes comparable to the free-fall time. The average typical density of these elementary cloudlets, called "clumpuscules", is $10^9 \, \text{cm}^{-3}$, column density $10^{24} \, \text{cm}^{-2}$, size 30 AU, and mass $10^{-3} \, \text{M}_\odot$. These small units are the building blocks of a fractal structure, that ranges upwards over 4 to 6 orders of magnitude in scales. They are gravitationally bound, and the corresponding thermal width along the line of sight for molecular hydrogen at $T = 3$ K is about $0.1 \, \text{km s}^{-1}$.

In Section 4 several problems raised by a cold gas dark matter are discussed, such as the stability and consistency of self-gravitating cold gas discs, the flatness of rotation curves, the effect of intense large scale magnetic fields, and implications at scales larger than galaxies.

Several of the hypotheses and conclusions in this and the companion papers have a speculative character. Only observations, particularly absorption line studies, will allow to test whether dark matter in galaxies can be, like in the solar neighbourhood, mostly or perhaps entirely baryonic, and, more specifically, in a form of hydrogen sufficiently diluted to permit star formation.

## 2. Constraints from Theory and Simulations

Starting in the late 60's, global dynamical studies of galaxies, in particular $N$-body simulations, have profoundly changed the representation we should have of galaxies. Nowadays people familiar with $N$-body simulations no longer consider galaxies as rigid objects formed for once $10-15 \, \text{Gyr}$ ago in a short time and remaining frozen until now, as proposed in the classical ELS scenario (Eggen et al. 1962). On the contrary, discs are viewed as fragile structures (e.g. Quinn 1987) since they respond strongly to typical perturbers even as light as the Magellanic Clouds. Interactions intermittently shake galaxies and can trigger important large-scale transformations. Tidal interactions have been extensively studied (e.g. Toomre & Toomre 1972; Barnes 1988; Hernquist 1990). On the observational side, evidences for a decreasing but non-negligible amount of interactions and mergers with distance to galaxy cluster centers are now much better documented (e.g. Whitmore 1993). Also many cosmological indications, such as the variation of quasar density with redshifts, or the Butcher-Oemler effect (Butcher & Oemler 1978), indicate that galaxies and their environment have changed significantly within a time-scale of 10 Gyr.

*2.1. Disc formation and instabilities*

First, the formation of discs by gravitational collapse in a turbulent gaseous medium is a process badly characterised by a unique time-scale (as assumed in the ELS scenario), but by a range of widely different time-scales. Since Larson's (1969) and other's simulations of gaseous gravitational collapses with non-negligible pressure, the initial free-fall time $\tau_{\text{ff}} = (G\rho_0)^{-1/2}$ is known to be the shortest time-scale of the collapse characterising only the central region. In an indefinitely extended medium, the collapse rapidly becomes inhomogeneous, pressure *gradients* develop, and the collapse proceeds at larger radii for arbitrary long times. Since dynamical processes are much faster in the inner parts than in the outer parts, the inner parts can already form a quasi-steady disc, while the outer parts still behave chaotically. For several tens of rotation periods the boundary conditions for the inner parts are not the ones of an isolated object: the concept of "formation epoch" is misleading. The slow growth of discs is vividly illustrated in high resolution cosmological $N$-body simulations with gas (e.g. Evrard 1992).

Second, dissipative discs are characterised by a wide choice of gravitational instabilities with relatively fast growth rates ($\approx 10\,\mathrm{Gyr}^{-1}$) (see e.g. Binney & Tremaine 1987, Chap. 6). Owing to the fact that angular momentum is less easily dissipated than energy, a cooling disc tends to minimise kinetic pressure while conserving circular motion. Inevitably at some point the Safronov-Toomre stability parameter $Q$ (Safronov 1960; Toomre 1964) has to become critical, i.e. the disc becomes gravitationally unstable, highly responsive to perturbations.

A surprising large number of spirals, particularly the gas-rich Sc's, look chaotic, strongly suggesting a non-equilibrium state of these galaxies. Several $N$-body experiments (Hohl 1971; Sellwood & Carlberg 1984) display also chaotic spiral structures when an inelastic gas component is included. The basic mechanism which appears to be able to regenerate the spiral structure in galaxies beyond a few revolutions is the so-called "swing amplification" (Toomre 1981). However such a process heats the discs rapidly to such an extent that it needs a constant supply of fresh cold gas to be maintained. Gas infall could be the solution of the problem of maintaining the chaotic spiral structure, as suggested by Toomre (1990). The privilege that only Sc's would receive gas seems strange unless earlier type galaxies would be former Sc's having evolved toward other types.

### 2.2. The bar-bulge connection

Typically, the evolution of a critically stable $N$-body disc is that it easily forms a bar and spiral arms, allowing then a fast global transfer of angular momentum. Therefore the classical assumption that galactic discs remain almost unchanged over more than 10 Gyr is also easily invalidated.

$N$-body simulations have shown early (Hohl 1971) that a bar reorganises a stellar disc into precisely the observed characteristic exponential profile in the outer region (Freeman 1970) with a steeper profile in the bar-bulge region. At that time Hohl's result was not taken seriously into account since bars were viewed as anomalies. But now they appear instead to be the rule (a lucid explanation why bars are more natural than perfect axisymmetric discs in the region with a linearly raising rotation curve was given by von Weizsäcker already in 1951, but forgotten during the following decades). More accurate simulations have confirmed Hohl's finding (Combes et. al 1990; Pfenniger & Friedli 1991). Therefore the dynamical effects of bars cannot be ignored.

Another finding of $N$-body work (Combes & Sanders 1981), also not taken seriously first (Sellwood & Wilkinson 1993), was that self-consistent bars excite important instabilities transverse to the galactic plane, tending to thicken the disc in $z$ and generating a peanut-shaped bulge aspect of the bar. Bars are sufficiently non-axisymmetric structures for widening significantly all the orbital resonances, increasing chaotic motions. In addition to the corotation, Lindblad and other radial resonances, bars have important vertical resonances able to amplify star oscillations about the galactic plane (Pfenniger 1984). The variety of numerically generated peanut-shaped bars is restricted, and their density profiles quantitatively resemble the observed light profiles of peanut-shaped galactic bulges such as NGC 128 (Friedli & Pfenniger 1990). The connection between bars and peanut-shaped bulges was also confirmed by independent, more accurate, and more numerous calculations (Combes et al. 1990; Raha et al. 1991).

Recent studies of secular effects in discs have shown that the accumulation of gas, or any kind of matter, within the bar region (more precisely within its inner Lindblad resonance) is able first, to produce a secondary smaller bar (Friedli & Martinet 1993), and second, to later dissolve the bar into a hot spheroid half as large as the bar ($\approx$ the disc scale-length) akin to a small bulge (Hasan & Norman 1990; Pfenniger & Norman 1990; Friedli & Pfenniger 1990). The required quantities of matter are only of the order of 2% of the galaxy stellar mass, illustrating the fragility of discs. In usual spirals such a mass transfer provoked by the gravity torques on the gas can be expected over a span of a few Gyr. So stellar bars appear to be fragile against dissipative perturbations and to be able to dissolve into small bulges especially fast in gas rich galaxies. A bar can eventually reform later if the disc cooling and sufficient gas accumulation bring the disc again to instability. After each bar dissolution the stellar mass of the bulge increases, since bulge building is an irreversible process.

An alternative possibility to form bulges is accretion of satellites. However, the formation of *small bulges* such as Sc's bulges by accretion is a very unlikely process, because it requires to dissipate the energy of accreting satellites with an atypical small amount of angular momentum *without heating the rest of the disc*. In other words all the satellite kinetic energy must be deposited near the centre at the first passage, because inevitably subsequent passages cross the disc and heat it. In fact, $N$-body simulations show that the repeated merging of dwarf satellite galaxies 1) destroys a bar (Pfenniger 1991a) and 2) thickens a disc into possibly big bulges: 10% of such a kind of mass accretion can produce *large bulges* like in the Sombrero galaxy (Pfenniger 1992, 1993b).

### 2.3. Energetics and disc sizes

Simple global energetics considerations (Pfenniger 1991b, 1992) show that the power emitted by stars in galaxies is comparable to the gravitational power $L_{\mathrm{gra}}$ that a galaxy can exchange by large scale dynamics ($L_{\mathrm{gra}} \approx V^5/G$ is the ratio of the gravitational energy $GM^2/R$ to the dynamical time $R/V$ of a virialised system $RV^2 = GM$, where $V$ is the virial velocity, $M$ the mass, $R$ the radius, and $G$ the gravitational constant). Thus, unless disc galaxies would be completely transparent, which are not since a substantial fraction of the light is thermalised and reemitted in the (far) infrared, on the long run stellar light represents a power sufficiently large to counteract gravitation, leading to the dimensional relation $L_{\rm bol} \sim V^5/G$ linking the bolometric luminosity $L_{\rm bol}$ to the gravitational power $L_{\rm gra}$. Comparing the actual luminosities and inserting the rotation velocities in this relation gives a striking match to the Tully-Fisher relation, in view of the invoked order of magnitude arguments. Note that the $L_{\rm bol} \sim V^5/G$ relation applies only to a completely self-regulated star forming disc. The lower exponent usually found in the IR Tully-Fisher relation can result from the contamination of young disc light by bulge and old disc light. To first order an old stellar population has a constant $L/M$ ratio, therefore it should follow closely the virial theorem: $L \sim V^2 R$. The observed correlation for the ellipticals, called the "fundamental plane", is indeed very close to this relation (Djorgovski 1994).

The power coincidence $L_{\rm bol} \approx L_{\rm gra}$ in actively star forming discs can only be understood if star formation is effectively self-regulated by a fast feed-back with global dynamics (Quirk 1972; Kennicutt 1989, 1990). It means that stellar activity[2] (internal physics), rather than initial conditions, determines galaxy sizes. Exactly the same is thought to be correct for stars, i.e. the relevant factor determining the size and mass of stars is not the initial fluctuation spectrum of the diffuse ISM, but the internal physics (mainly the atomic properties of hydrogen).

If star formation in discs is self-regulated with the disc global dynamics, then it implies again that the aspect of disc galaxies must change over a time-scale comparable with the gas consumption time-scale, often found much shorter than the Hubble time with the conventionally determined amount of gas.

*2.4. Evolution along the Hubble sequence*

From the above considerations a more definite picture of the secular evolution of galaxies emerges. Disc galaxies can form from large scale and slowly rotating primordial gas condensations as poorly structured, bulgeless and mostly gaseous discs, such as Im, Sm galaxies. As gas cools rapidly by radiation, proto-galaxies tend to flatten and to become more axisymmetric. Stellar activity is the principal energetic factor controlling the galaxy sizes.

The virial theorem tells us that the energy loss increases the rotation speed simultaneously to the binding energy: slowly rotating Sd's are energetically less evolved than fast rotating Sa's. The square of the typical rotation speed ratios $((100\,{\rm km\,s^{-1}}/350\,{\rm km\,s^{-1}})^2 \approx 1/12)$ indicates the ratios of mechanical energy dissipation. The increasing number of rotations progressively symmetrises the outer disc, and winds up the spiral arms. Effectively Sd's are typically asymmetric, Sc's have patchy and well open spiral arms, while Sa's-S0's are almost symmetric and have tightly wounded spiral arms.

General cooling brings discs towards gravitational instability, leading to the formation of a bar, that secularly forms a bulge about half as large as the bar, so a small bulge. Beyond a critical accumulation of mass near the centre, bars dissolve into a spheroidal shape. Big bulges larger than the disc scale-length can result from the merging of a few dwarf satellite galaxies. So, secularly the mass of the bulge increases. Bulge growth is indeed an irreversible process; the Hubble sequence, from Sd to S0, is well known to be also a sequence of increasing bulge/disc ratio.

The other irreversible process indicating the sense of evolution along the Hubble sequence is star formation (Kennicutt 1990, 1992) since star formation continuously consumes gas and blocks most of it in long lived low mass stars. The consequence of nucleosynthesis, of stellar winds, and of supernovae is to increase the amount of dust and heavy elements.

Therefore disc galaxies may only evolve along the Hubble sequence from Im–Sm–Sd to Sa–S0. The typical time-scale for a major change in a galaxy, such as the transformation from one Hubble type to the next, can be as short as 1 Gyr for dynamical factors and also of the order of 1 Gyr for star formation factors. But this time-scale is highly variable, since fortuitous merger events of arbitrary amplitude can boost not only dynamical evolution, but can trigger bursts of star formation.

*2.5. M/L ratio and transformation of dark matter into stars*

All these considerations about galaxy evolution have a direct implication for the nature of dark matter. If really galaxies do evolve along the Hubble sequence in the sense Sd to Sa, with a typical time-scale shorter than 10 Gyr, we have to understand why the ratio dark mass to stellar mass decreases systematically from Im–Sm–Sd to Sa–S0 by a factor $10^2$ while the ratio dark mass to HI mass remains constant, around $10 - 30$ (Broeils 1992).

Also the approximate conspiracy connected with rotation curves has to be contemplated in time. Detailed studies of the age-metallicity and age-velocity relations in the solar neighbourhood have concluded towards a nearly constant star formation rate (SFR) in the local disc during its evolution (Carlberg et al. 1985; Meusinger et al. 1991). A nearly steady state in SFR is also found in extragalactic studies (Kennicutt 1983, 1992; Donas et al. 1987). However Larson et al. (1980) had noticed that the average time-scale for consuming all the disc gas by star formation is of the order of 4 Gyr, in contradiction with the absence of evolution of the SFR; this problem is known as the "gas consumption problem" (cf. also Tinsley 1980). In fact, the constant SFR concerns mainly the middle spirals (from Sb to Sc) but not the later types and irregulars, affected by bursts (Scalo 1988), or the earlier types, that appear to have a declining SFR accompanying their gas depletion. If the SFR is indeed steady in a large isolated spiral, this could be obtained by the gradual extension of the diameter of its optical disc into regions with fresh gas, which are also dominated by dark matter. The mysterious conspiracy of flat rotation curves follows then naturally.

---

[2] Here "stellar activity", not to confuse with "chromospheric activity", is meant to include all the phenomena linked with star formation and evolution, and with all the mass, momentum and energy exchanges occurring between the stars and the ISM.

Often fresh gas infall is invoked to explain the persistence of star formation (e.g. Larson et al. 1980). But discs are fragile; the amount of mass falling in from arbitrary directions and with arbitrary speeds is restricted for galaxies such as the Milky Way to a few percent of mass in order to conserve the flatness of discs (Quinn 1987; Tóth & Ostriker 1992), a much too small amount for compensating the average rate of star formation.

Preliminary calculations of the gas consumption time, quoted by Kennicutt (1992), indicate that the gas consumption problem can be less severe than previously estimated when the recycling of the gas is not assumed to be instantaneous, and a higher star mass loss is adopted. Delayed gas recycling and large stellar winds extend the star formation time-scale by a factor two to four, still shorter than the Hubble time. So the question remains to understand how the fraction of dark matter can be less in early type galaxies than in late type galaxies, while presumably early type galaxies were late type galaxies $5-10$ Gyr ago.

These evolutionary considerations and apparent problems suggest that galaxy evolution would seem much simpler if *dark matter were in a form able to generate later stars*, i.e. fresh diluted hydrogen and helium gas. This hypothesis excludes not only exotic particles, but also dense objects such as brown or white dwarfs, Jupiters, or black-holes.

Since along the Hubble sequence the estimated amount of HI is systematically of the order of 1/10 of the dark mass, while the ratio of star mass to gas mass varies by two orders of magnitudes, it is clear that if a systematic error in the gas mass determination would be as large as a factor $10-30$, essentially no dark matter would be required in discs (Casertano & van Albada 1990).

Therefore, we are led to the idea that more gas than previously thought, sufficiently diluted in order to be able to form stars later on, should be available in the outer discs. This gas would be mostly supported by rotation, shrinking in radius only slowly due to the difficulty to dissipate angular momentum. The energy dissipation has to be small since outer HI discs seem to persist for time-scales much longer than the HI cooling time without apparent corresponding energy input. In paper II a solution is proposed in which a very clumpy (fractal) structure of cold gas ($T \approx 3$ K) with quasi-isothermal conditions can considerably decrease the energy dissipation rate. Star formation, destroying the isothermal conditions in the optical disc would increase the dissipation rate at the transition from the optical to the HI disc. Gaseous dark matter would be consumed by star formation. The most likely possibility is that both gas would accrete slowly inwards, because dissipation cannot strictly vanishes, and the optical disc would spread out with time, because the optical disc, made of dissipationless stars, can hardly shrink in radius. The exact amount of both inward gas motion and outward spreading of the stellar "fire" cannot be presently well estimated theoretically owing to our rudimentary knowledge of the involved physics of the ISM, star formation, and galactic dynamics.

## 3. Observational Constraints on the Dark Gas

We review in this Section the constraints that observations impose on the possible forms of mainly hydrogen.

*3.1. Halo gas in the Milky Way*

3.1.1. Halo gas in the solar neighbourhood

The halo gas in our own Galaxy has been surveyed through optical and ultraviolet absorption studies in front of halo stars (e.g. Edgar & Savage 1989; Danly et al. 1992), and in front of remote quasars or X-ray emitters (e.g. Morton & Blades 1986; Kinney et al. 1991). An important result from UV studies is that the low ionisation material is about a factor of 10 times more abundant than the high ionisation material (e.g. Danly 1991). From the comparison of 21 cm HI emission towards the same line of sight as the absorbed halo stars, the $z$-distribution of halo gas can be derived. Danly et al. found evidence in most lines of sight for gas beyond $z = 1$ kpc, with column densities of the order of $1-12 \cdot 10^{19}$ atoms cm$^{-2}$. If the gas layer is fitted by an exponential $z$-profile, the exponential scale-length is 510 pc. There is an excess of negative velocities for the high-$z$ gas seen in absorption, which has supported speculations on galactic fountains as the origin of this gas (e.g. Shapiro & Field 1976; Shapiro 1991). No evidence of neutral gas above the lower halo do exist: no further absorption lines occur towards extragalactic sources with respect to high latitude halo stars (above 2–3 kpc).

Diffuse ionised gas has been also studied through pulsar dispersion measures. Since millisecond pulsars have been recently discovered in globular clusters at high $z$ ($\sim 4$ kpc), the $z$-distribution of the ionised medium has been derived (Reynolds 1989). A characteristic scale-height of 1.5 kpc has been found for free electrons, significantly greater than that from neutral gas studies (Dickey & Lockman 1990). Highly ionised gas, traced by absorption and emission lines in UV, provides also evidence for a large exponential scale-height, roughly 3 kpc (Savage & Massa 1987). The volume filling factor of this component is low, between 3 and 10%. There is some indication for a substantial increase of the highly ionised species (Si IV, C IV) above 1 kpc (Pettini & West 1982), suggesting an extragalactic ionisation source.

Above 1 kpc, the gaseous medium is therefore dominated by the diffuse ionised gas, that could be in transition regions between the HI clouds and the hot coronal gas. The source of ionisation, either from young stars or supernovae remnants from the plane (McKee & Ostriker 1977), or from the extragalactic EUV background (York 1982; Heisler & Ostriker 1988) is still subject of debate. The observed amount of NV appears to require the existence of collisionally ionised gas near $2 \cdot 10^5$ K (Savage et al. 1990).

Molecules have also been detected at high latitudes, in emission (Magnani et al. 1985), and absorption (de Vries & van

Dishoeck 1988; Welty et al. 1989), but never at $z$ larger than 250 pc. Martin et al. (1990) have detected the $H_2$ fluorescence in the Lyman band towards the diffuse interstellar medium. They detect molecular hydrogen at high latitude, even when CO remains undetected because photodissociated, and suggest the presence of molecular halos, resulting from the photodissociation of dense clumps.

We can deduce from the Milky Way emission and absorption studies that there is indeed evidence for cool gas at high latitude; however, the possibility of disc-halo interactions (galactic fountains, bubbles, chimneys, etc.) prevents us to derive column densities for the primordial halo gas. Also, if the dark gas around galaxies is a rotationally supported component, more in the shape of a thick disc than in the shape of a non-rotating spherical component, then we are at a unfavourable place in the Milky Way to detect such a component; external galaxies should give better constraints.

### 3.1.2. Global disc and halo gas

Only $\gamma$-ray measurements can inform us about the total number of nucleons integrated along the line of sight. They result from the interaction between cosmic ray protons and matter. The Galaxy is virtually transparent to $\gamma$-rays (Bloemen 1989), but the gas distribution is not known independently from the cosmic ray (CR) distribution. Therefore, no exact estimate of the total gas amount can be derived. In fact, the exponential scale-length of the $\gamma$-ray radial distribution is 15 kpc, much larger than that ($\sim 3-5$ kpc) of the potential CR sources such as supernovae, pulsars, and massive stars, or also main sequence F to M stars which may contribute significantly to CR by sunspot activity (Cassé & Goret 1978; Meyer 1985). This is interpreted in the frame of CR diffusion models (e.g. Bloemen 1989), but could also trace some hidden form of hydrogen.

Indeed, first CR isotopic abundances studies show that the bulk of cosmic rays near the Sun comes from a distance of about 3 kpc from the galactic centre (Maeder & Meynet 1993). Second, CR particles are closely attached to the interstellar magnetic field lines, because of their small gyration radius, and the magnetic field intensity is a function of the volume density. The large radial extension of $\gamma$-rays point towards a larger volume density of gas. At large distances however, the gas volume density falls off at least as $R^{-2}$, implying also a fall-off of CR. Most of the outer dark gas could then not be traced by $\gamma$-rays since the amount of CR sources is low there.

### 3.2. Gas around external galaxies

### 3.2.1. Neutral gas

In the 21 cm emission line neutral gas is detected in most galaxies out to $1-2R_H$ at a level of about $10^{20}$cm$^{-2}$, where $R_H$ is the Holmberg radius (e.g. Bosma 1981; Sancisi 1988). Recent high-sensitivity observations have reached currently levels of $10^{18}$cm$^{-2}$, without increasing the HI sizes of galaxies. Van Gorkom et al. (1993) have undertaken a deep VLA survey of the galaxy NGC 3198 to test the HI extent at a resolution of 1 arcmin (3 kpc), and discovered a considerably sharp decline of HI emission, suggesting the existence of an ionising front. The HI emission extends only exceptionally beyond $2R_H$ (Huchtmeier & Richter 1982; Carignan & Freeman 1988). Many of these systems have an extended HI distribution due to a tidal origin (Weliachew et al. 1978; Simkin et al. 1987). An interesting feature in the outer parts of galaxies is the HI shoulder noticed by Sancisi (1988): after the optical disc radius, the radial distribution of HI surface density changes its slope (e.g. Wevers et al. 1986). The radial distribution varies then inversely proportional to radius, as the dark matter (Bosma 1981). The radial HI extent is also considerably reduced for galaxies in clusters, especially in the center of X-ray emitting clusters (Cayatte et al. 1990). This HI-depletion is attributed to ram pressure stripping by the hot intracluster gas.

### 3.2.2. Molecular gas

The molecular gas radial extent is unknown in spiral galaxies. Only the CO emission yields indirect information on the $H_2$ content, and in general CO is observed only within the optical disc (e.g. Young & Scoville 1991, and references therein). In our own galaxy, weak CO emission has been detected at large radii (Wouterloot et al. 1990; Digel et al. 1993), somewhat outside the optical disc, but radial distances are hard to measure outside the solar circle. Recently, Allen & Lequeux (1993) observed also CO absorption in front of extragalactic continuum sources, at large radii. However, the CO/$H_2$ conversion ratio depends on the metallicity, and radial gradients of abundances are well known (Pagel & Edmunds 1981). Moreover, the surface density of the gas is rapidly decreasing with radius. If there exists a diffuse phase, the CO molecule looses its self-shielding well before molecular hydrogen, and could be entirely photodissociated, as in diffuse clouds in the solar vicinity (Blitz et al. 1990). Therefore, and essentially for metallicity reasons, the absence of detected CO emission does not preclude the existence of large amounts of $H_2$.

### 3.2.3. Diffuse ionised gas

Diffuse ionised gas could be traced in external galaxies by $H\alpha$ emission. For example Dettmar (1990) discovered a large extension (several kpc) of diffuse ionised gas perpendicular to the plane of NGC 891. He derives an electronic typical scale-height of 600 pc. However, this large $z$-extent is restricted to the inner half of the optical disc, and is *highly related to star formation*. Rand et al. (1990) founds an extended filamentary structure, with an even larger scale-height, between 2 and 3 kpc. The ionised gas layer is therefore thicker than what is obtained for the Milky Way by Reynolds (1989). This could be related to the higher star formation rate in NGC 891. The high-$z$ $H\alpha$ emission reveals a cut-off at $R = 5-6$ kpc. In the outer parts, $H\alpha$ observations are not sensitive enough to detect column densities as high as $10^{21}$ cm$^{-2}$, spread over a line of sight of 3 kpc, and the corresponding constraints on ionised gas around spiral galaxies are not significant (Braine & Combes 1993).

## 3.3. More molecular gas in interacting galaxies and starbursts

Recently, Braine & Combes (1992, 1993) have shown that interacting galaxies, showing signs of tidal perturbations, have stronger and more centrally concentrated CO emission than unperturbed systems. If the CO/H$_2$ conversion ratio is roughly constant, according to the non-varying molecular line ratios, then this implies a larger amount of molecular gas in interacting galaxies, by a factor 4–5 with respect to unperturbed ones. This larger amount is normalised to the blue luminosity, or to the physical size of the systems. The HI gas, on the contrary, is observed in about the same amount in interacting galaxies, with respect to unperturbed ones. Only for the stronger interactions and mergers is the HI content lower.

Braine & Combes (1993) suggested a scenario in which large hydrogen reservoirs exist in the outer parts of galaxies, in rotational equilibrium. The strong gravity torques induced by an interacting companion, could drive progressively the gas towards the center (Combes 1991b). Gas infall could be more violent and sudden during the merging of equal-mass galaxies (Barnes & Hernquist 1991). Some evidence for such a sudden infall of "primordial" gas is found in the isotopic line ratios of molecular emission in mergers (Casoli et al. 1992).

Fresh gas infall from the outer parts of galaxies helps to understand the existence of huge starbursts in mergers, and their exceptional infrared luminosities. In particular, a molecular mass of more than $10^{11}$ M$_\odot$ has recently been discovered in a distant infrared galaxy, of a total $L_{IR} = 10^{14}$ L$_\odot$ (Rowan-Robinson et al. 1991; Brown & van den Bout 1991; Solomon et al. 1992). The fact that the HI content begins to decrease in violent mergers, may suggest that they are running out of gas, or that the infall rate is insufficient.

## 3.4. QSO absorption lines

Maybe the clearer evidence for the existence of a large amount of gas around spiral galaxies is the widespread detection of absorption lines in front of quasars. The large frequency of Ly-$\alpha$ absorptions (up to 100 systems in a single line of sight), remains unexplained. The derived mean cross-section for Ly-$\alpha$ is huge: assuming that they originate in galaxies puts the effective radius of spiral galaxies to 480 kpc, at a mean redshift of $z = 2.5$ (Sargent 1988). Also metal-enriched low-ionisation systems with $N$(HI) larger than $10^{18}$ cm$^{-2}$ are found with large sizes, about 3–4 $R_H$ (Bergeron & Stasinska 1986; Bergeron & Boisse 1991). Cases of 21 cm HI line absorbers of remote quasars by high-redshift systems have been reported (e.g. Briggs 1988). The derived column densities are of the order of $10^{21}$ cm$^{-2}$. Two low redshift absorbers have been identified with nearby galaxies (Haschick & Burke 1975; Carilli & van Gorkom 1987), and heavy elements are also present (Boksenberg & Sargent 1978; Rubin et al. 1982). However, Carilli et al. (1989) demonstrated that at least in one of these cases, the extended HI is due to a tidal tail, and proposed that most low redshift absorption line systems belong to interacting galaxies.

Absorption systems with damped Ly-$\alpha$ lines are also characterised by their high column densities (larger than $10^{20}$ cm$^{-2}$), and are thought to be associated to galactic discs (Wolfe 1988). Their occurrence frequency along the line of sight of quasars is larger than normal spiral galaxies; this implies that spiral discs at that epoch ($z \sim 2$) had column densities of $2 \cdot 10^{20}$ cm$^{-2}$ up to 3.5 $R_H$, instead of 1.2 $R_H$ measured in nearby spirals in HI emission, assuming the same comoving density of disc galaxies.

As for molecular hydrogen, only one case, at $z = 2.8$, of H$_2$ absorption has been reported (Levshakov & Varshalovich 1985; Foltz et al. 1988) with an estimated column density of $\sim 10^{18}$ cm$^{-2}$. Observations are however difficult to perform and to interpret, given the multiplicity of lines in the Lyman and Werner bands, and the confusion with the Ly-$\alpha$ forest.

Recently the Ly-$\alpha$ forest has been detected at low redshift with the Hubble Space Telescope towards 3C273 and PKS 2155-304 (Morris et al. 1991; Bruhweiler et al. 1993). The large number of lines detected (9–16) was a surprise; the Ly-$\alpha$ absorbers being much more frequent than expected by extrapolation of the high-redshift evolution. The number of lines was even compatible with no evolution since $z = 2$. According to their metallicity and clustering characteristics, the absorbers appear to have the same basic properties as those at high redshift. *The abundance of gas around galaxies is therefore maintained at low redshift*. An alternative interpretation has been recently proposed in terms of ionised gas (Maloney 1992), extending a few hundreds of kpc around galaxies.

## 3.5. Assumptions leading to neutral gas mass

When the hydrogen mass is determined from the 21 cm line observations, several assumptions are made, the violation of each one leading to an underestimate of the mass. Recent extensive reviews about HI observations have been written by Kulkarni & Heiles (1987) and Burton (1992).

In galactic conditions the 21 cm line emission is almost always collisionally excited (for a density larger than about $5 \cdot 10^{-3}$ cm$^{-3}$). The actually measured signal is the difference between the brightness temperature and the background temperature, which is proportional to the difference between the gas kinetic and background radiation temperatures. Often the background temperature is neglected, but it is at least at 3 K. So any gas cold enough to have a temperature close to the background temperature is barely detectable, independent of the optical depth. Hence one should be careful before ruling out very cold conditions especially because the cold gas cooling time by black-body radiation is always very short with respect to the Hubble time. In fact, observations of CO and continuum radiation at 20 cm in the Galaxy and nearby galaxies suggest that if the main cold gas heating source is cosmic rays, then large amounts of unseen cold gas near 3 K should exist (Allen 1993).

A very important assumption usually made is that HI is optically thin. Extremely thin column densities below about $10^{18}$ cm$^{-2}$ are difficult to detect, and for temperatures below 100 K, column densities higher than about $10^{21}$ cm$^{-2}$ are opti-cally thick. Therefore the range of measurable column densities in HI is practically less than $10^3$, and more typically of the order of $10^2$. Similar restrictions apply to other single line observations (cf. Scalo 1990). Larger ranges of column densities can be investigated only by a series of lines. A priori the ISM possesses a range of densities between the intergalactic medium density ($< 10^{-3}\,\mathrm{H\,cm^{-3}}$) and the stellar density ($\sim 10^{24}\,\mathrm{H\,cm^{-3}}$). In comparison, the highest measured densities below the ones of stars are of the order of $10^7 - 10^9\,\mathrm{H\,cm^{-3}}$ only (i.e. Mezger et al. 1988; Migenes et al. 1989; Wilking et al. 1989). These are condensations associated with star formation, i.e. warm cores. In some cases they correspond to accretion discs around stars.

Observational surveys show that HI in the Galaxy can be both very transparent, since one observes well at large distances throughout the disc, and completely opaque, since the brightness toward the center and anti-center of the Galaxy are equal (as explained by e.g. Burton 1992, Chap. 2.4). Such seemingly contradictory conclusions are only possible if the HI distribution is extremely clumpy. Despite the Doppler shift that makes HI more transparent to the 21 cm line, in several other directions at low Galactic latitudes the 21 cm optical depth hovers around $\tau = 1$ (Burton 1992). *Therefore the assumption of general optical thinness for the* 21 cm *line is not supported by observations, particularly because* HI *is known to be clumpy up to the highest resolution*. The main argument for optical thinness, that the 21 cm line can be detected at large distances in the Galactic disc, is invalid if HI is far from being homogeneous, actually the general evidence.

As for the molecular phase traced by the CO emission, it is well known that the CO line is generally optically thick; the $CO/H_2$ conversion factor is empirical, justified by the virial theorem (e.g. Combes 1991a). This ratio is unlikely to be valid when the metallicity is far from solar, or when the excitation is insufficient (Allen & Lequeux 1993).

### 3.6. Formation of molecular hydrogen

Molecular hydrogen is the lowest energy state of neutral hydrogen. Naively one should therefore expect its frequent occurrence in space. Dissociating UV radiation and slow chemical reaction rate in a low density medium around $1\,\mathrm{cm^{-3}}$ are usually invoked to rule out large fractions of molecular hydrogen in the outer disc.

Both assumptions can be vitiated in a hierarchical medium that both shields itself and contains dense clumps, and is remote from massive star formation sites such as in the outer galactic discs. Indeed the time-scale of $H_2$ formation by simple neutral-neutral reaction, $3 \times 10^6/n_H$ Gyr (see e.g. Genzel 1992) is much longer than the Hubble time in a homogeneous medium with $n_H \approx 1\,\mathrm{cm^{-3}}$, but *becomes short with respect to galaxy ages in high density clumps with $n_H > 10^5\,\mathrm{cm^{-3}}$*. If ionising radiation or dust are present in some small amount, catalytic reactions can form $H_2$ much faster than the neutral-neutral reaction. Therefore the above $H_2$ formation time-scale is a conservative upper bound.

Such high density clumps are common in molecular clouds where molecules with heavy elements allow us to observe them, but molecules such as CO just trace mass and are unlikely the unique cause of these high densities. No $CO/H_2$ relation can obviously exist in a high density pure H+He medium in which no stars have ever enriched it with heavy elements.

### 3.7. Fractal geometry of cold gas

The fractal nature of the interstellar cold gas has been increasingly well documented (Falgarone et al. 1992). Nearby molecular clouds are observed self-similar in projection over a range of scales and densities of at least $10^4$, but perhaps up to $10^6$.

According to measures of HI in VLBI and of variable emission of QSO's, structures on a scale as small as 25 AU (Diamond et al. 1989), or 10 AU (Fiedler et al. 1987) have been detected. Churchwell et al. (1987) also have detected 22 compact radio sources in Orion, whose sizes range from less than 27 AU to 230 AU. Cometary globules in the Helix planetary nebula (Meaburn et al. 1992) are examples of dense ($> 6 \cdot 10^5\,\mathrm{H\,cm^{-3}}$) and small ($\approx 60\,\mathrm{AU}$) clumps that seem to be evaporated by the central hot star.

Remarkably the fractal dimension of cold gas *isophotes* at different wavelengths is almost constant, around $1.3 - 1.4$, over a wide range of scales, and independent of the molecular or atomic state of the gas, or whether it is thought to be self-gravitating or not (Scalo 1990; Falgarone 1992).

Modelling the physics of fractals is presently far from trivial, as one of the most basic traditional assumptions in physics, differentiability, is lost in practice, due to the huge dynamical range needed to describe the strong gradients. Unless the smallest scale of a hierarchical structure is accessible (to observations, respectively to calculations), where differentiability might be a valid assumption, in principle the use of differential equations is not justified if the resolution is insufficient. *In particular the usual radiation transfer equation and the hydrodynamical description need a differentiable medium*.

Often physical properties can greatly differ between smooth and fractal objects, because macroscopic properties may depend on the smallest scale at which the fractal behaviour stops. Although the general evidence is that HI is structured down to the smallest accessible scales, lacking better tools observers apply the equation of transfer and assume homogeneity at subresolution scales to estimate the amount of interstellar gas, which in turn provides the amount of baryons in the outer discs of spirals.

In a companion paper (Paper II), and in Pfenniger 1993a, we show by fractal models that large underestimates of the mass can result by factors more than 10 when assuming that the fractal dimension is 3 (diffuse medium) while in reality it would be less, around $1.5 - 2$.

### 3.8. Summary

Only a small fraction of the mass can exist in a diffuse, partially ionised warm gas which would then correspond to the metallic lines of ionised systems observed in absorption in front of quasars. This phase is to be expected, since the outer parts of galaxies are reached by UV and soft X-ray photons from quasars (Heisler & Ostriker 1988; Hasinger et al. 1991). Possibly the sharp HI emission fall-off in the outer parts observed by van Gorkom et al. (1993) is due to the ionisation of this diffuse phase. Since the diffuse phase fills most of the volume at a low density, it cannot contribute significantly to dark matter.

On the other hand the existence of a cold, dense, self-shielded and very low volume-filling gas in fractal form is compatible with observations. This is the only form where a large amount of baryonic mass can be contained. A factor 10 or more of hydrogen mass underestimate is enough to remove the need of exotic matter in disc galaxies (cf. Casertano & van Albada 1990). In addition to large optical depths, both the formation of $H_2$ and a low temperature close to 3 K may considerably increase the amount of mass not emitting the 21 cm line.

## 4. Discussion

The subject of dark matter, even when restricted to spirals, corresponds to such an extended literature that all its aspects cannot be discussed in a single paper. There has been good reviews on the dark matter problem with different emphases, by Trimble (1987), Carr (1990), Ashman (1992), and Tremaine (1992). We briefly mention below some points that seem relevant to us.

### 4.1. Disc stability and self-consistency

A question that arises immediately if disc dark matter is assumed to be made of cold gas is to explain the stability of such discs when the radial velocity dispersion is inferred from the observed HI or CO velocity dispersion ($\sigma_{\text{gas}} = 5 - 14 \, \text{km s}^{-1}$). Linked to this is the problem of explaining how dark matter might be confined to a disc as thin as the flaring HI discs when considering the low velocity dispersion of the gas.

In order to prevent gravitational radial instabilities the Safronov-Toomre stability parameter

$$Q = \frac{\kappa \sigma_{\text{gas}}}{\pi G \Sigma} \, , \qquad (1)$$

(Safronov 1960; Toomre 1964; Goldreich & Lynden-Bell 1965) must satisfy $Q \gtrsim 1$. As usual $\kappa$ is the radial epicyclic frequency and $\Sigma$ is the disc surface density. This criterion is often evaluated with the Mestel's thin disc model $\Sigma \sim R^{-1}$. In a such thin disc with flat rotation curve $v_c = $ constant, we have $\kappa = \sqrt{2}v_c/R$ and $\Sigma = v_c^2/2\pi G R$ (cf. Binney & Tremaine 1987, Chap. 2.6). With $v_c = 200 \, \text{km s}^{-1}$ stability requires $\sigma_{\text{gas}} \approx Qv_c/\sqrt{8} \gtrsim 71 \, \text{km s}^{-1}$, independent of radius.

In outer parts of disc galaxies $\sigma_{\text{gas}}$ is indeed observed to be almost constant (e.g. NGC 1058 by Dickey et al. 1990), which is consistent with a $Q$-regulated gaseous disc with a radial density profile proportional to dark matter (consistent with the Bosma HI/dark matter ratio). However with respect to 71 km s$^{-1}$, $\sigma_{\text{gas}}$ is too small by a factor $\approx 10$, which is usually taken as an argument for invoking an external thick dark matter halo.

Actually observed HI discs appear as far from being perfectly axisymmetric and smooth; they can present large asymmetries, warps, spiral arms and massive HI complexes, so the necessity of strict stability is not demonstrated (see e.g. M33 and M101 as reviewed by Brinks 1990, or even the prototypical dark matter spiral NGC 3198, Begeman 1989). The irregular shapes of HI discs from large scale down to the beam resolution suggest on the contrary that the gas is unsteady, and a subcritical $Q$ might be acceptable. Also when a disc is substantially non-axisymmetric, the applicability of Eq. (1) is less obvious. Instead one might consider average quantities over several turns, or over $2\pi$ in azimuth at a given time. Then the effective radial velocity dispersion is increased by the contribution of the radial non-circular motions[3]. To first order for a constant circular velocity $v_c$, in the average disc the epicyclic velocity is $\approx \kappa \Delta R = \sqrt{2}v_c \Delta R/R$, where $\Delta R$ is the non-axisymmetric offset at the radius $R$. The radial component of the epicyclic velocity is smaller by a factor $1/\sqrt{2}$. So the effective radial velocity dispersion is given by

$$\sigma_{\text{eff}}^2 \approx \left(\frac{\Delta R}{R}\right)^2 v_c^2 + \sigma_{\text{gas}}^2 \, , \qquad (2)$$

For example for $\Delta R/R = 1/10$, $v_c = 200 \, \text{km s}^{-1}$ and $\sigma_{\text{gas}} = 10 \, \text{km s}^{-1}$, we get $\sigma_{\text{eff}} \approx 22 \, \text{km s}^{-1}$.

Furthermore the HI discs are far from being thin. On the contrary they often flare out approximately linearly with radius beyond the optical disc, before being warped at larger distances (see e.g. Burton 1992, Merrifield 1992), giving thus the impression of an exponential flaring. The general effect of a finite thickness on the stability is to increase the effective $Q$ by a factor $1 + kh_z$ (Vandervoort 1970, Romeo 1992), where $k$ is the wave-number and $h_z$ the disc scale-height, because mass being diluted in $z$ the responsiveness of the disc is decreased. If we replace $\Sigma$ in Eq. (1) by $2\rho(R)h_z$, were $\rho(R)$ is the mass density in the plane, and for flat rotation curves we have also $\rho(R) = \rho(R_0)R_0^2/R^2$, we obtain

$$h_z = \frac{\sqrt{2}}{2\pi G} \left(\frac{\sigma_{\text{gas}} v_c}{Q\rho(R_0)R_0^2}\right) R \, . \qquad (3)$$

So if the ratio within parentheses is constant, $h_z$ increases indeed linearly with radius, which is precisely the behaviour observed by Merrifield (1992) in the Milky-Way for $R_0 \lesssim R \lesssim 2R_0$ (see his Fig. [7]). Let us estimate $Q$ for the Galaxy with data from Merrifield's paper: $\sigma_{\text{gas}} = 10 \, \text{km s}^{-1}$, $v_c = 200 \, \text{km s}^{-1}$, $\rho(R_0) = 0.1 \, M_\odot \, \text{pc}^{-3}$, $R_0 = 8 \, \text{kpc}$, and his Fig. [7] gives $h_z \approx 0.045 \, R$. Thus from Eq. (3) we extract $Q \approx 0.36$. If a part of the observed flaring is contaminated by the warp, the true $h_z/R$ would be smaller and $Q$ would be increased. If taking into account non-axisymmetric motions $\sigma_{\text{eff}} = 20 - 30 \, \text{km s}^{-1}$, we reach a domain of subcritical $Q \sim 0.8 - 0.9$. Finally, if we take the finite thickness into account, the lowest wave-number $k$ being $2\pi/R$, we must multiply $Q$ by a factor of at least $1 + 2\pi h_z/R \approx 1.28$, which yields a marginally stable disc, $Q_{\rm eff} \approx 1$.

---

[3] It is easy to show that the continuity and momentum equations have the same forms for time-averaged quantities, except that the pressure term is increased by the square of the time-average velocity variations. It follows that the same stability criterion can be applied for the average quantities.

Let us check whether the above data is consistent with a completely self-gravitating disc as thin as the observed Milky-Way gaseous disc. For flat rotation curves we have $dK_z(z)/dz = 4\pi G\rho(z)$, where $K_z$ is the force along $z$. Integrating the Jeans equation[4]

$$\rho\sigma_z^2(z) = \int_z^\infty \rho K_z \, dz = \frac{1}{4\pi G} \int_z^\infty \frac{dK_z}{dz} K_z \, dz \,, \quad (4)$$

gives $\sigma_z^2(0) = K_z(\infty)^2/8\pi G\rho(0)$. Approximating $dK_z/dz(0)$ by $K_z(\infty)/h_z$ gives $h_z = \sigma_z(0)/\sqrt{2\pi G\rho(0)}$. Self-consistency with flat rotation curves requires $\rho = \rho(R_0)(R_0/R)^2$, which yields to

$$h_z = \frac{\sigma_z(0)}{\sqrt{2\pi G\rho(R_0)}} \frac{R}{R_0} \,. \quad (5)$$

Again $h_z$ is proportional to $R$ if $\sigma_z$ is constant. Assuming $\sigma_z(0) = \sqrt{2}\,\sigma_{\rm gas}$ (averaged over $z$, $\sigma_z$ is certainly smaller than $\sigma_z(0)$) we get, using the previous numerical values, $h_z = 0.034R$, so quite close to the observed (and probably warp-contaminated) $h_z = 0.045R$ taken from Merrifield's paper.

Let us discuss the possible bias due to the effect of an outer disc flaring and a truncation at a finite radius on the usual mass estimate by the rotation curve. Indeed in a truncated Mestel disc the rotation speed increases *inside* and near the truncation radius, which obviously results in an overestimate of the inner mass. In several disc mass models with either constant height or linearly increasing height, we have compared, by numerical quadratures, the effective inner mass with the mass deduced from the virial relation $GM = v_c^2 R$. In order to conserve an almost flat rotation curve some mass models have been taken as $\rho \sim \arccos(R/R_{\rm max})/R^2$, where $R_{\rm max}$ is the truncation radius. The general result is that at $R < R_{\rm max}$ the inner mass is overestimated, with respect to the virial value, from $\approx 0\%$ at $R = 0$ up to $\lesssim 30-40\%$ at $R = R_{\rm max}$. An overestimate by 40% is certainly an upper bound for realistic discs with a mild truncation. The effect of flaring has been found negligible. *Thus the disc truncation effect can only decrease the required dark matter by a small amount.*

Therefore taking into account the non-axisymmetric aspect of HI discs and their finite and flaring thickness alleviates the need of large $\sigma_{\rm gas}$. A $\sigma_{\rm gas}$ of the order of $\sim 10\,{\rm km\,s}^{-1}$, appears much less a problem in a realistic non-axisymmetric HI disc than in the perfect Mestel disc. Also linearly flaring and completely self-gravitating gas disc are possible with a velocity dispersion of the order of only $10\,{\rm km\,s}^{-1}$.

---

[4] A reasonable assumption is that $K_z$ tends asymptotically towards a constant or a maximum at a $z$ where $\rho$ is negligible. Here the symbol $\infty$ is understood as the height at which $K_z$ reaches the maximum.

In fact the problem of stability is not restricted to large scale. Cold gas in outer discs is strongly Jeans unstable already at small scales ($\lesssim 100\,{\rm pc}$), as clearly visible in the Milky Way by the profusion of details in the HI (e.g. Burton 1992), yet it survives there for Gyr without apparent energy input. In Paper II we argue that the paradox of cold gas persistence without formation of dense objects such as Jupiters or stars can be understood by its fractal structure at a temperature close to 3 K. At the smallest scale of the fractal hierarchy the smallest clumps, called clumpuscules, are prevented to collapse by collisions because the collision time is much smaller than the collapse time. The smallest scale of the fractal is determined by the transition from the isothermal to the adiabatic regime. As a consequence of collisions the clumpuscules exchange energy with their neighbours and with the 3 K radiation. The near isothermality suppresses energy dissipation. The effect of the inclusion of the 3 K radiation for the energetics of outer discs does not seem to have been investigated. Indeed, in the outer discs the energy density of the 3 K radiation ($u = 4\sigma T^4/c \approx 0.26\,{\rm eV\,cm}^{-3}$) becomes comparable to the turbulent kinetic energy of the cold gas beyond $R \approx 20\,{\rm kpc}$.

### 4.2. Flat rotation curves and global energy minimum

Traditionally the flatness of the rotation curves is explained by the presence of a massive and hot collisionless isothermal dark halo at the kinetic temperature of the rotation speed. Now, if dark matter is in a cold gaseous form and self-gravitating, one should understand why the rotation curves tend to be flat.

If most of the mass is cold, it has to be supported mostly by rotation and the condition of isothermality no longer leads a priori to constant rotation curves. However, the total energy $E$ of a self-gravitating and rapidly rotating disc is easily seen to be a functional $E[\Phi]$ of the potential $\Phi$ alone

$$E[\Phi] = \iint 2\pi R \, dR \, dz \, \left(\tfrac{1}{2}\rho v_c^2 + \tfrac{1}{2}\rho\Phi\right) \,,$$
$$\text{where} \quad \rho = \frac{\nabla\Phi}{4\pi G}, \quad \text{and} \quad v_c^2 = R\frac{\partial\Phi}{\partial R} \,. \quad (6)$$

Minimising $E[\Phi]$ by variational calculus restricts the class of potential functions $\Phi$ precisely to the ones having a flat rotation curve (Pfenniger 1989):

$$\delta E[\Phi] = 0 \quad \text{requires} \quad \frac{\partial v_c^2}{\partial R} = 0 \,. \quad (7)$$

This result is obtained without constraining the boundary conditions, just what is needed for slowly growing discs with time-dependent and chaotic outskirts. This result is robust in the sense that small perturbations to the total energy integral in Eq. (6) leads to small perturbations of the result in Eq. (7). Also perturbations produced by mass distributions outside the considered volume do not change the result. In any case, flat rotation curves are incompatible with boundary conditions of isolated objects, since mass diverges proportionally to the radius.

12  D. Pfenniger et al.: Is dark matter in spiral galaxies cold gas? I.*4.3. Magnetic fields*

Magnetic fields are sometimes invoked for solving the dark matter problem in discs (e.g. Nelson 1988; Battaner et al. 1992), with the implicit hope that by complicating the physics new alternatives can emerge.

Yet an elementary argument[5] based on the virial theorem shows that any magnetic field *increases* the dark matter problem. In the virial theorem (cf. Chandrasekhar 1968, p. 581) the different interacting energies of a self-gravitating systems in uniform expansion or contraction (and *in particular* in equilibrium), sum up to zero. Except the gravitational energy, in an isolated system all the conceivable energies contribute to the sum positively, including the magnetic energy.

Admittedly, if the system is non-isolated the magnetic (or gas) pressure at the boundary contributes negatively to the virial equation. But this negative contribution can outweight its internal positive contribution only if the outer pressure is larger than the volume-averaged internal pressure. If they are equal, their effect cancel.

Therefore increasing the magnetic field intensity inside a galaxy implies, that at the admitted known rotation speed and size of the galaxy *more mass* is required to compensate the additional magnetic pressure. The observed rotation velocity implies a *lower bound* on dark mass. Additional non-circular motions, gas pressure or magnetic fields increase the positive terms in the virial equation and therefore increase the required mass.

Finally, if a continuous conducting gas can eventually be supported by magnetic fields at the galactic scale, this assumption is harder to justify in a highly hierarchical and turbulent gas in which densities vary by more than five orders of magnitude[6]. In any case young stellar populations born from the gas certainly cannot be supported by a general magnetic field at the galactic scale. Therefore if gas would be supported mainly by magnetic fields and would rotate much faster than gravitation would require alone, then the orbital angular momentum of young stars would exceed the one of circular orbits, and, contrary to observations, the velocity dispersion of young stellar populations would be much larger than the velocity dispersion of the gas.

Other arguments against a significant rôle of magnetic fields have been presented by Persic & Salucci (1993): the kinematics of galaxy pairs, satellite galaxies, and small groups all require dark halos, not magnetic fields.

*4.4. Polar rings and warps*

As for galaxy companions, polar rings are indicators of the local gravitational potential, so they constraint the total inner mass, but they weakly constraint its distribution.

---

[5] During the refereeing phase of this paper very similar arguments have been published in Nat. 356 (1993) by J.R. Jokipii & E.H. Levy (p. 19), and by P. Cuddeford & J. Binney (p. 20).
[6] In comparison the density difference between common gas and solid objects on Earth spans only 3–4 decades.

Over the years polar ring studies have less and less constrained the dark halo to be spherical. Recent works (Arnaboldi 1992) indicate the possibility of relatively flat ($c/a \approx 0.6$) halos. We remark that constraining the shape of the mass distribution by the shape and kinematics of the orbits of test particles is an intrinsically difficult task. Major modifications of the galaxy mass distribution far from the polar ring change the orbit shapes very little, while, on the contrary, the self-gravity of a polar ring as light as a few percents of the total mass can not only determine its stability, but also modify the orbit shape. This is a crucial point when the polar ring is considered as a test body in the potential of a dark halo (Sparke 1986).

Concerning warps, it is interesting to recall an often forgotten paper by Petrou (1980). She showed that precisely a dark halo that becomes more flattened with radius may be a solution to the problem of explaining why observed warps have a straight line of nodes. A priori the warp line of nodes should rapidly wind up in a differentially rotating disc. In the context of spherical, or slightly triaxial massive halos no convincing solution to the line of nodes problem has been proposed yet.

A related argument for a flat dark halo in the outer discs has been put forward by Sancisi & van Albada (1987): If the outer gas in spirals may have been accreted recently, as suggested by its asymmetric and irregular shape, the fact that it has already almost found the optical disc plane indicates the existence of a massive flat disc in the outer parts, more or less coplanar with the optical disc.

*4.5. Dark matter at larger scales*

Kinematical studies concerning the galaxy rotation curves in clusters are still based on too small samples to infer general conclusions on the presence, or absence, of an important mass in the outer part of galaxies in such environments. More and better resolved optical as well as 21 cm data are necessary.

A clear HI deficiency is observed for the galaxies in the inner parts of the clusters with respect to the more external regions, generally attributed to a stripping process. Whitmore et al. (1988) claim that in these galaxies optical rotation curves drop somewhat in the regions beyond the HI limit, while Amram et al. (1993) contest the reality of the declining rotation curves. It is too early to say if there is a correlation between the HI deficiency and a decrease of the rotation curve.

At the scale of clusters, more dark matter appears to be present, with a density parameter in the range $\Omega = 0.1 - 0.3$ (e.g. Peebles 1984). The possibility for this matter to be baryonic too is not excluded. Although this remains to be examined in details, some recent discoveries about hot gas in clusters makes it attractive. Mushotzky (1993) presented X-ray observations of galaxy clusters in which the hot gas of large clusters is able to represent a mass comparable to the dark mass.

The distribution of dark matter inferred from the observations of gravitational images in the shape of arcs is clearly more peaked in the center of the cluster than the visible matter (Hammer 1991). This is also confirmed by the interpretation of the hot gas X-ray emission (Gerbal et al. 1992). In rich clusters, where most of the galaxies are of early type, so are not surrounded by a lot of cold dark gas, most of the gas must be in a common envelope, with a characteristic velocity dispersion which can be, at some scales, lower than the intra-cluster dispersion.

Nucleosynthesis constraints permit the possibility of baryonic dark matter, since the baryonic mass contribution to the necessary mass closing the Universe, $\Omega_b$ in the frame of the standard Big Bang nucleosynthesis could be as high as $\Omega_b = 0.1$ if the Hubble constant is within its lowest possible values (e.g. Boesgaard & Steigman 1985; Smith et al. 1993). However other uncertainties already in the frame of this model could raise $\Omega_b$ to 0.2 (Schramm 1991). In the frame of inhomogeneous cosmologies $\Omega_b$ can reach 1 (e.g. Malaney & Fowler 1988). The visible baryons already account for $\Omega_b \approx 0.01$, and we propose that around galaxies, the dark baryons increase this contribution to $\Omega_b \approx 0.1$. In any case, the nucleosynthesis lower limit already requires a lot of dark baryons.

## 5. Conclusions

We have recalled that the arguments for the presence of dark matter in spirals have become weaker for the optical parts of galaxies, while dark matter is really needed in outer gaseous discs. Other pieces of information now suggest that disc galaxies evolve faster than thought 30 years ago. In a couple of Gyr, a bar can form and dissolve, rearranging the disc into an exponential profile, and heating the central disc vertically into a bulge. Already internal factors such as star formation and large scale dynamical instabilities speed up evolution, but external perturbations, such as satellite interactions and mergers, trigger also large scale instabilities and star formation. The rapid transformation of gas into stars, when contemplated with the dark to stellar mass conspiracy in the rotation curves, particularly points to look for the possible identification between dark matter and a form of matter able to form stars, i.e. sufficiently diluted gas.

Better observations of the ISM show that the cold gas is fractal and essentially clumpy down to very small scales, of the order of a few tens of AU. A wide range of gas densities ($\gtrsim 10^9\,\mathrm{H\,cm^{-3}}$) that continuity between dense clumps and stars leads to infer the existence, are difficult to investigate by today's techniques. As shown in Paper II, HI mass determination could underestimate the gas mass by a factor 10 or more owing to the very inhomogeneous nature of cold gas leading to high optical depths in a small fraction of the sky, rapid $H_2$ formation, and thermal equilibrium with the 3 K radiation. The dark matter to HI constant ratio would seem then natural. If such a large error in the gas mass determination can be confirmed by new observations, then the problem of dark matter in discs would be solved.

*Acknowledgements.* We are grateful to John Scalo for refereeing thoroughly this work. His criticisms allowed us to improve significantly the presentation of the paper. This work was supported partly by the Swiss National Science Foundation.

## References

Allen R.J., 1993, in: Back to the Galaxy, S.S. Holt, F. Verter (eds.), AIP, New York, p. 287
Allen R.J., Lequeux J., 1993, A&A 410, L15
Amram P., Sullivan III S.T., Balkowski C., Marcellin M., Cayatte V., 1993, ApJ 403, L59
Arnaboldi M., 1992, Dynamics and Evolution of Polar Rings Galaxies, PhD Thesis, SISSA, Trieste
Ashman K., 1992, PASP 104, 1109
Athanassoula E., Sellwood J.A., 1986, MNRAS 221, 213
Bahcall J.N., Casertano S., 1985, ApJ 293, L7
Bahcall J.N., Flynn C., Gould A., 1992, ApJ 389, 234
Barnes J., 1988, ApJ 331, 699
Barnes J., Hernquist L., 1991, ApJ 370, L65
Battaner E., Garrido J.L., Membrado M., Florido E., 1992, Nat 360, 652
Begeman K.G., 1989, A&A 223, 47
Bergeron J., Boisse P., 1991, A&A 243, 344
Bergeron J., Stasinska G., 1986, A&A 169, 1
Bienaymé O., Robin A., Crézé M., 1987, A&A 180, 94
Binney J., Tremaine S., 1987, Galactic Dynamics, Princeton Univ. Press, Princeton
Blitz L., Bazell D., Désert F.X., 1990, ApJ 352, L13
Blitz L., Binney J., Lo K.Y., Bally J., Ho P.T.P, 1993, Nat 361, 417
Bloemen H., 1989, ARA&A 27, 469
Boesgaard A.M., Steigman G., 1985, ARA&A 23, 319
Boksenberg A., Sargent W.L., 1978, ApJ 220, 42
Bosma A., 1981, AJ 86, 1971
Bottema R., 1993, A&A 275, 16
Braine J., Combes F., 1992, A&A 264, 433
Braine J., Combes F., 1993, A&A 269, 7
Briggs F.H., 1988, in: QSO absorption lines: Probing the Universe, Blades J.C., Turnshek D.A., Norman C.A. (eds.), Cambridge Univ. Press, Cambridge, p. 275
Brinks E., 1990, in: The Interstellar Medium in Galaxies, H.A. Thronson, J.M. Shull (eds.), Kluwer, Dordrecht, p. 39
Broeils A., 1992, Dark and visible matter in spiral galaxies, PhD Thesis, Rijksuniversiteit Groningen
Brown R.L., van den Bout P.A., 1991, AJ 102, 1956
Bruhweiler F.C., Boggess A., Norman D.J., Grady C.A., Urry M.C., Kondo Y., 1993, ApJ 409, 199
Burton W.B., 1992, in: The Galactic Interstellar Medium, Saas-Fee Advanced Course 21, D. Pfenniger, P. Bartholdi (eds.), Springer-Verlag, Berlin, p. 1
Butcher H., Oemler A., 1978, ApJ 219, 18
Carignan C., Freeman K.C., 1988, ApJ 332, L33
Carilli C.L., van Gorkom J.H., 1987, ApJ 319, 683
Carilli C.L., van Gorkom J.H., Stocke J.T., 1989, Nat 338, 134
Carr B.J., 1990, Comments Astrophys. 14, 257
Carlberg R.G., Dawson P.C., Hsu T., van den Berg D.A., 1985, ApJ 294, 674
Casertano S., van Albada T.J., 1990, in: Baryonic Dark Matter, D. Lynden-Bell, G. Gilmore (eds.), Kluwer, Dordrecht, p. 159

Casertano S., van Gorkom J.H., 1991, AJ 101, 1231
Casoli F., Dupraz C., Combes F., 1992, A&A 264, 55
Cassé M., Goret P., 1978, ApJ 221, 703
Cayatte V., van Gorkom J.H., Balkowski C., Kotanyi C., 1990, AJ 100, 604
Chandrasekhar S., 1968, Hydrodynamic and Hydromagnetic Stability, Oxford Univ. Press, London
Churchwell E., Felli M., Wood D.O.S., Massi M., 1987, ApJ 321, 516
Combes F., 1991a, in: Dynamics of Galaxies and their Molecular Cloud Distributions, F. Combes, F. Casoli (eds.), IAU Symp. 146, Kluwer, Dordrecht, p. 255
Combes F., 1991b, ARA&A 29, 195
Combes F., Debbasch F., Friedli D., Pfenniger D., 1990, A&A 233, 82
Combes F., Sanders R.H., 1981, A&A 96, 164
Danly L., 1991, in: The Interstellar Disk-Halo Connection in Galaxies, IAU Symp. 144, H. Bloemen (ed.), Kluwer, Dordrecht, p. 53
Danly L., Lockman F.J., Meade M.R., Savage B.D., 1992, ApJS 81, 125
Dettmar R.-J., 1990, A&A 232, L15
de Vaucouleurs G., 1963, ApJS 8, 31
de Vries C.P., van Dishoeck E.F., 1988, A&A 203, L23
Diamond P.J., et al., 1989, ApJ 347, 302
Dickey J.M., Hanson M.M., Helou G., 1990, ApJ 352, 522
Dickey J.M., Lockman F.J., 1990, ARA&A 28, 215
Digel S., de Geus E., Thaddeus P., 1993, ApJ, preprint
Djorgovski S., 1994, in: Ergodic Concepts in Stellar Dynamics, V.G. Gurzadyan, D. Pfenniger (eds.), Springer, in press
Donas J., Deharveng J.M., Laget M., Milliard B., Huguenin D., 1987, A&A 180, 12
Edgar R.J., Savage B.D., 1989, ApJ 340, 762
Eggen O.J., Lynden-Bell D., Sandage A.R., 1962, ApJ 136, 768
Einasto J., Kaasik A., Saar E., 1974, Nat 250, 309
Evrard A.E., 1992, in: Physics of Nearby Galaxies, Nature or Nurture?, T.X. Thuan, C. Balkowski, J.T.T. Van (eds.), Editions Frontières, Gif-sur-Yvette, p. 375
Falgarone E., 1992, in: Astrochemistry of Cosmic Phenomena, P.D. Singh (ed.), Kluwer, Dordrecht, p. 159
Falgarone E., Puget J.-L., Pérault M., 1992, A&A 257, 715
Fiedler R.L., Dennison B., Johnston K.J., Hewish A., 1987, Nat 326, 675
Foltz C.B., Chaffee F.H., Black J.H., 1988, ApJ 324, 267
Freeman K.C., 1970, ApJ 160, 811
Freeman K.C., 1992, in: Physics of Nearby Galaxies, Nature or Nurture?, T.X. Thuan, C. Balkowski, J.T.T. Van (eds.), Editions Frontières, Gif-sur-Yvette, p. 201
Friedli D., Benz W., 1993, A&A 268, 65
Friedli D., Martinet L., 1993, A&A 277, 27
Friedli, D., Pfenniger, D., 1990, in: Bulges of Galaxies, ESO Conf. and Workshop Proc. 35, B.J. Jarvis, D.M. Terndrup (eds.), p. 265
Genzel R., 1992, in: The Galactic Interstellar Medium, Saas-Fee Advanced Course 21, D. Pfenniger, P. Bartholdi (eds.), Springer-Verlag, Berlin, p. 275

Gerbal D., Durret F., Lima-Neto G., Lachièze-Rey M., 1992, A&A 253, 77
Goldreich P., Lynden-Bell D., 1965, MNRAS 130, 125
Hammer F., 1991, ApJ 383, 66
Hasan H., Norman C., 1990, ApJ 361, 69
Haschick A.D., Burke B.F., 1975, ApJ 200, L137
Hasinger G., Schmidt M., Trümper J., 1991, A&A 246, L2
Heisler J., Ostriker J.P., 1988, ApJ 332, 543
Hernquist L., 1990, in: Dynamics and Interactions of Galaxies, R. Wielen (ed.), Springer-Verlag, Heidelberg, p. 108
Hohl F., 1971, ApJ 168, 343
Hohl F., Hockney R.W., 1969, J. Comput. Phys. 4, 306
Huchtmeier W.K., Richter O.G., 1982, A&A 109, 331
Kalnajs A.J., 1987, in: Dark Matter in the Universe, IAU Symp. 117, J. Kormendy, G.R. Knapp (eds.), Reidel, Dordrecht, p. 289
Kennicutt R.C., 1983, ApJ 272, 54
Kennicutt R.C., 1989, ApJ 344, 685
Kennicutt R.C., 1990, in: Evolution of the Universe of Galaxies, R.G. Kron (ed.), ASP Conf. Ser. 10, p. 141
Kennicutt R.C., 1992, in: Star Formation in Stellar Systems, G. Tenorio-Tagle, M. Prieto, F. Sánchez (eds.) Cambridge Univ. Press, Cambridge, p. 193
Kinney A.L., Bohlin R.C., Blades J.C., York D.G., 1991, ApJS 75, 645
Kuijken K., Gilmore G., 1989, MNRAS 239, 571, & 605
Kulkarni S.R., Heiles C., 1987, in: Interstellar Processes, D.J. Hollenbach, H.A. Thronson (eds.), Reidel, Dordrecht, p. 87
Kundić T., Hernquist L., Gunn J.E., 1993, in: Back to the Galaxy, S.S. Holt, F. Verter (eds.), AIP, New York, p. 592
Larson R.B., 1969, MNRAS 145, 271
Larson R.B., Tinsley B.M., Caldwell C.N., 1980, ApJ 237, 692
Levshakov S.A., Varshalovich D.A., 1985, MNRAS 212, 517
Lewis J.R., Freeman K.C., 1989, ApJ 97, 139
Lin C.C., Shu F.H., 1964, ApJ 140, 646
Maeder A., Meynet G., 1993, A&A 278, 403
Magnani L., Blitz L., Mundy L., 1985, ApJ 295, 402
Malaney R.A., Fowler W.A., 1988, ApJ 333, 14
Maloney P., 1992, ApJ 398, L89
Martin C., Hurwitz M., Bowyer S., 1990 ApJ 354, 220
McKee C.F., Ostriker J.P., 1977, ApJ 218, 148
Meaburn J., Walsh J.R., Clegg R.E.S., Walton N.A., Taylor D., Berry D.S., 1992, MNRAS 255, 177
Meusinger H., Reimann H.-G., Stecklum B., 1991, A&A 245, 57
Merrifield M.R., 1992, AJ 103, 1552
Meyer J.P., 1985, ApJS 57, 173
Mezger P.G., Chini R., Kreysa E., Wink J.E., Salter C.J., 1988, A&A 191, 44
Migenes V., Johnston K.J., Pauls J.A., Wilson T.L., 1989, ApJ 347, 294
Miller R.H., Prendergast K.H., 1968, ApJ 151, 699
Morris S.L., Weymann R.J., Savage B.D., Gilliland R.L., 1991, ApJ 377, L21
Morton D.C., Blades J.C., 1986, MNRAS 220, 927

Mushotzky R.F., 1993, in Relativistic Astrophysics and Particle Cosmology, ed. C.W. Akerloff, M.A. Srednicki, Ann. N.Y. Acad. Sci., 688, 184
Nelson A.H., 1988, MNRAS 233, 115
Oort J.H., 1932, Bull. Astron. Inst. Netherlands 6, 249
Ostriker J.P., Peebles P.J.E., 1973, ApJ 186, 467
Pagel B.E.J., Edmunds M.G., 1981, ARA&A 19, 77
Peebles P.J.E., 1984, ApJ 284, 439
Persic M., Salucci P., 1993, MNRAS 261, L21
Petrou M., 1980, MNRAS 191, 767
Pettini M., West K.A., 1982, ApJ 260, 561
Pfenniger D., 1984, A&A 134, 373
Pfenniger D., 1989, ApJ 343, 142
Pfenniger D., 1991a, in: Dynamics of Disc Galaxies, B. Sundelius (ed.), Göteborg University, Göteborg, p. 191
Pfenniger D., 1991b, in: Dynamics of Disc Galaxies, B. Sundelius (ed.), Göteborg University, Göteborg, p. 389
Pfenniger D., 1992, in: Physics of Nearby Galaxies, Nature or Nurture?, T.X. Thuan, C. Balkowski, J.T.T. Van (eds.), Editions Frontières, Gif-sur-Yvette, p. 519
Pfenniger D., 1993a, in: Back to the Galaxy, S.S. Holt, F. Verter (eds.), AIP, New York, p. 584
Pfenniger D., 1993b, in: Galactic Bulges, IAU Symp. 153, H. Dejonghe, H. Habing (eds.), Kluwer, Dordrecht, in press
Pfenniger D., Combes F., 1994, A&A in press (Paper II)
Pfenniger D., Friedli D., 1991, A&A 252, 75
Pfenniger D., Norman C.A., 1990, ApJ 363, 391
Quinn P.J., 1987, in: Nearly Normal Galaxies, S.M. Faber (ed.), Springer, New-York, p. 138
Quirk W.J., 1972, ApJ 176, L9
Raha N., Sellwood J.A., James R.A., Kahn F.D., 1991, Nat 352, 411
Rand R.J., Kulkarni S.R., Hester J.J., 1990, ApJ 352, L1
Rees M.J., 1987, in: Dark Matter in the Universe, IAU Symp. 117, J. Kormendy, G.R. Knapp (eds.), Reidel, Dordrecht, p. 395
Reynolds R.J., 1989, ApJ 339, L29
Roberts M.S., Rots A.H., 1973, A&A 26, 483
Rogstad D.H., Shostak G.S., 1972, AJ 176, 315
Romeo A.B., 1992, MNRAS 256, 320
Rowan-Robinson M., Broadhurst T., Lawrence A. et al., 1991, Nat 351, 719
Rubin V.C., Thonnard N., Ford W.K., 1982, AJ 87, 477
Safronov V.S., 1960, Annales d'Astrophysique 23, 979
Salucci P., Ashman K.M., Persic M., 1991, ApJ 379, 89
Sancisi R., 1988, in: QSO absorption lines: Probing the Universe, Blades J.C., Turnshek D.A., Norman C.A. (eds.), Cambridge Univ. Press, Cambridge, p. 241
Sancisi R., van Albada T.S., 1987, in: Dark Matter in the Universe, IAU Symp. 117, J. Kormendy, G.R. Knapp (eds.), Reidel, Dordrecht, p. 67
Sargent W.L.W., 1988, in: QSO absorption lines, J. Blades, D. Turnshek, C. Norman (eds.), Cambridge University Press, p. 1
Savage B.D., Massa D., 1987, ApJ 314, 380
Savage B.D., Massa D., Sembach K., 1990, ApJ 355, 114
Scalo J., 1988, in: Evolution of Galaxies, J. Palouš (ed.), Prague Astron. Inst., p. 101
Scalo J., 1990, in: Physical Processes in Fragmentation and Star Formation, R. Capuzzo-Dolcetta et al. (eds.), Kluwer, Dordrecht, p. 151
Schramm D.N., 1991, in: After the First Three Minutes, S.S. Holt, C.L. Bennett, V. Trimble (eds.), AIP, New York, p. 12
Sellwood J.A., Carlberg R.G., 1984, ApJ 282, 61
Sellwood J.A., Wilkinson A., 1993, Rep. Prog. Phys. 56, 173
Shapiro P.R., 1991, in: The Interstellar Disk-Halo Connection in Galaxies, IAU Symp. 144, H. Bloemen (ed.), Kluwer, Dordrecht, p. 417
Shapiro P.R., Field G.B., 1976, ApJ 205, 762
Simkin S.M., van Gorkom J., Hibbard J., Su H.-J., 1987, Science 235, 1367
Smith M.S., Kawano L.H., Malaney R.A., 1993, ApJS 85, 219
Solomon P.M., Downes D., Radford S.J.E., 1992, ApJ 398, L29
Sparke L.S., 1986, MNRAS 219, 657
Tinsley B.M., 1980, Fundamentals of Cosmic Physics 5, 287
Tinsley B.M., 1981, MNRAS 194, 63
Toomre A., 1964, ApJ 139, 1217
Toomre A., 1981, in: The Structure and Evolution of Normal Galaxies, S.M. Fall, D. Lynden-Bell (eds.), Cambridge Univ. Press, p. 111
Toomre A., 1990, in: Dynamics and Interactions of Galaxies, R. Wielen (ed.), Springer-Verlag, Heidelberg, p. 292
Toomre A., Toomre J., 1972, ApJ 178, 623
Tóth G., Ostriker J.P., 1992, ApJ 389, 5
Tremaine S., 1992, Phys. Today 45, 28
Trimble V., 1987, ARA&A 25, 425
van Albada T.S., Sancisi R., 1986, Philos. Trans. R. Soc. London A 320, 447
Vandervoort P.O., 1970, ApJ 161, 87
van Gorkom J.H., Cornwell T., Sancisi R., van Albada T.S., 1993, preprint
von Weizsäcker C.F., 1951, ApJ 114, 165
Weliachew L., Sancisi R., Guelin M., 1978, A&A 65, 37
Welty D.E., Hobbs L.M., Blitz L., Penprase B.E., 1989, ApJ 346, 232
Wevers B.M.H.R., van der Kruit P.C., Allen R.J., 1986, A&AS 66, 505
Whitmore B.C., Forbes D.A., Rubin V.C, 1988, ApJ 333, 542
Whitmore B.C., 1993, in: Physics of Nearby Galaxies, Nature or Nurture?, T.X. Thuan, C. Balkowski, J.T.T. Van (eds.), Editions Frontières, Gif-sur-Yvette, p. 351, 425
Wilking B.A., Mundy L.G., Blackwell J.H., Howe J.E., 1989, ApJ 345, 257
Wolfe A.M., 1988, in: QSO absorption lines: Probing the Universe, Blades J.C., Turnshek D.A., Norman C.A. (eds.), Cambridge Univ. Press, Cambridge, p. 297
Wouterloot J.G.A., Brand J., Burton W.B., Kwee K.K., 1990, A&A 230, 21
York D.G., 1982, ARA&A 20, 221
Young J.S., Scoville N.Z., 1991, ARA&A 29, 581
Zwicky F., 1933, Helvetica Physica Acta 6, 110